\documentclass[aps,pra,preprint,a4paper]{revtex4-1}
\usepackage{amsmath}
\usepackage{amssymb}
\usepackage{graphicx}
\usepackage{subfigure}

\newcommand{\bra}[1]{\ensuremath{\left\langle #1 \right|}}
\newcommand{\ket}[1]{\ensuremath{\left| #1 \right\rangle}}
\newcommand{\braket}[2]{\ensuremath{\left\langle #1 | #2 \right\rangle}}
\newcommand{\figref}[1]{Fig. \ref{#1}}

\begin{document}
\title{Noise enhanced performance of adiabatic quantum computing by lifting degeneracies}
\author{R.D. Wilson} 
\email{R.D.Wilson@lboro.ac.uk}
\author{A.M. Zagoskin} 
\author{S. Savel'ev}
\affiliation{Department of Physics, Loughborough University, Loughborough, Leicestershire LE11 3TU, UK}

\begin{abstract}
	We investigate the symmetry breaking role of noise in adiabatic quantum computing using the example of the CNOT gate. In particular, we analyse situations where the choice of initial Hamiltonian produces symmetries in the Hamiltonian and degeneracies in the spectrum. We show that, in these situations, the conventional stipulation that the initial and problem Hamiltonians do not commute is unnecessary as noise will inherently play the role of a universal symmetry breaking perturbation and split any level crossings that may impede or obstruct the computation. The effects of an artificial noise source with a tailored time-dependent amplitude are also explored and it is found that such a scheme could offer a considerable performance enhancement. These results are found using a novel, generalised version of the Pechukas-Yukawa model of eigenvalue dynamics.
\end{abstract}
\maketitle
\section{Introduction}

  Recently there has been a lot of interest in alternative paradigms to the standard approach to quantum computing (i.e. the quantum circuit model). Adiabatic quantum computing (AQC) is a promising example which is particularly suited to solving optimisation problems \cite{farhi-2000}. AQC involves slow adiabatic evolution from a configuration with an easily reachable ground state to one where the ground state encodes the solution to the hard computational problem in hand. This scheme is believed to have a number of advantages over the ``standard'' approach, namely, the precise time-dependent control of individual qubits is no longer required, and it benefits from an inherent robustness against some of the effects of decoherence by remaining in the instantaneous ground state at all times \cite{childs-2001, ashhab-2006, roland-2005}. Crucially, AQC has been shown to be polynomially equivalent, under certain conditions, to the standard gate model of quantum computing \cite{aharonov-2004, mizel-2006}.  The effects of noise on AQC are generally considered to be detrimental but manageable \cite{childs-2001, ashhab-2006, roland-2005}. It was nevertheless stated that its effect can increase the success probability of AQC in some situations by; providing an alternative evolution trajectory \cite{childs-2001}, or by thermal relaxation back to the ground state \cite{amin-2008}. Here, we investigate a more general effect of noise on AQC.Namely, how noise inherently breaks any hidden symmetries in the Hamiltonian and, thus, splits any level crossings in the energy spectrum which could impede or even prevent the computation.
	
	AQC can be described by the following Hamiltonian;
\begin{equation}
	\mathcal{H}(\lambda(t))=H_{0}+\lambda(t)ZH_{b},
	\label{eqn:basicHamiltonian}
\end{equation}
where the ground state of the final Hamiltonian, $H_{0}$, encodes the desired solution, $ZH_{b}$ is a large bias term (with $Z\gg 1$) and the initial configuration, $\mathcal{H}(\lambda =1)=H_{0}+ZH_{b}$, has an easily reachable, non-degenerate ground state. In order for there to be a high probability of the system remaining in the ground state as the bias is switched off, the rate of change of the control parameter $\lambda (t)$ must be sufficiently slow to suppress excitation via Landau-Zener-St\"{u}ckelberg tunneling \cite{landau-1932,zener-1932,stueckelberg-1932}.
  
  In AQC, the initial and final Hamiltonians are usually chosen such that no symmetries exist in $\mathcal{H}(\lambda(t))$ by ensuring that they do not commute. This is done to ensure that there are no degeneracies in the energy spectrum during the evolution. However, this restriction on the choice of initial Hamiltonian may prove practically impossible to realise for generic problems. In \cite{zagoskin-2007} the example of the adiabatic equivalent of the CNOT gate is discussed as an example of a prototypical quantum algorithm. It was shown that, the choice of a generic $H_b$ could lead to an abundance of level crossings in the energy spectrum for the $\ket{00}\rightarrow\ket{00}$ operation of the CNOT gate. The influence of these level-crossings on the performance of the adiabatic quantum algorithm was not considered there. However, the effects of level crossings in the spectrum were discussed in \cite{farhi-2000} and it was noted that the addition of an appropriate sort of perturbation to the system will break the symmetries and therefore split any level crossings. The question of how to provide an appropriate perturbation still remains open.

  In this paper, we propose that noise will inherently fulfil the role of the crucial symmetry breaking perturbation in physical implementations of an AQC system and as a result of this, the condition that $H_0$ and $H_b$ do not commute is no longer required. The CNOT gate algorithm is again used as an example of a prototypical quantum algorithm and we show that the performance of this generic algorithm is relatively resistant to the effects of noise, in agreement with \cite{childs-2001, ashhab-2006, roland-2005}. As reported in \cite{childs-2001}, we find that in some situations the presence of noise increases the success probability, we then go on to explore the relationship between this increase in success probability and the fidelity of the final state. We also discuss the idea of using a tailored artificial noise signal to try and enhance the performance of the adiabatic quantum computation process. To do this, we derive and utilise a generalised stochastic version of the Pechukas-Yukawa equations \cite{pechukas-1983, yukawa-1985}; where the dynamics of the energy eigenvalues are mapped exactly on to the classical dynamics of a 1D gas of Brownian particles with a mutual repulsive force.
  

 \section{Generalised Pechukas-Yukawa model} 
  
  The standard Pechukas-Yukawa model is derived from a Hamiltonian of the form \eqref{eqn:basicHamiltonian}. However, to incorporate a source of noise in to the model we start with the following Hamiltonian;
\begin{equation}
	\mathcal{H}(\lambda(t))=\mathcal{H}_{0}+\lambda(t)ZH_{b}+\delta h(\lambda(t)),
	\label{eqn:Hamiltonian}
\end{equation}
where the perturbation strength $\lambda(t)$ plays the role of `time' and the new stochastic term $\delta h(\lambda)$ describes random fluctuations in the Hamiltonian due to an external noise source. The instantaneous eigenvalues and eigenfunctions of \eqref{eqn:Hamiltonian} are denoted $x_{n}(\lambda)$ and \ket{n(\lambda)} respectively; $\mathcal{H}(\lambda)\ket{n(\lambda)}=x_{n}(\lambda)\ket{n(\lambda)}$. By following the same procedure as the derivation of the standard Pechukas-Yukawa model, as detailed in \cite{stockman-2006}, we arrive at the following generalised system of equations;
\begin{align}
	\dot{x}_{n}=\frac{\partial x_{n}}{\partial \lambda}=&v_{n}+\dot{\delta h}_{nn},\notag \\
	\dot{v}_{n}=\frac{\partial v_{n}}{\partial \lambda}=&\sum_{k\neq n}\left[\frac{2\left|l_{nk}\right|^{2}}{(x_{n}-x_{k})^{3}} + \frac{l_{nk}\dot{\delta h}_{kn}-\dot{\delta h}_{nk}l_{kn}}{(x_{n}-x_{k})^{2}}\right], \label{eqn:extPYsystem} \\
	\dot{l}_{nm}=\frac{\partial l_{nm}}{\partial \lambda}=&\sum_{k\neq m,n}\left[l_{nk}l_{km}\left(\frac{1}{(x_{n}-x_{k})^2} - \frac{1}{(x_{m}-x_{k})^2}\right)\right. \notag \\
	&+ \left.\frac{(x_{n}-x_{m})(l_{nk}\dot{\delta h}_{km}-l_{km}\dot{\delta h}_{nk})}{(x_{m} -x_{k})(x_{n} -x_{k})}\right] \notag \\
	& +\dot{\delta h}_{nm}(v_{m}-v_{n}) + \frac{l_{nm}(\dot{\delta h}_{nn}-\dot{\delta h}_{mm})}{(x_{n}-x_{m})}. \notag
\end{align}
where $v_{n}(\lambda)=\bra{n}ZH_{b}\ket{n}$ and $l_{nm}=(x_{n}-x_{m})\bra{n}ZH_{b}\ket{m}$ for $n\neq m$. These generalised equations can be used to describe a wider array of physical systems because of the inclusion of the $\delta h(\lambda)$ term without any additional assumptions or approximations.  Note that if the noise term $\delta h(\lambda(t))$ is identically zero the system of equations \eqref{eqn:extPYsystem} simply reduces to the standard Pechukas-Yukawa equations. The equations \eqref{eqn:extPYsystem} describe the dynamics of the energy eigenvalues of the Hamiltonian \eqref{eqn:Hamiltonian}, but also correspond to the classical dynamics of a 1D interacting gas, where the $n$th particle has position $x_{n}(\lambda)$ and velocity $v_{n}(\lambda)$ and the strength of the inter-particle force between the $n$th and the $m$th particles is described by $l_{nm}(\lambda)$. 

  In order to close the system of equations \eqref{eqn:extPYsystem}, we need to consider the nature of the noise term $\delta h(\lambda)$. In general, noise in any physical system arises from a number of independent sources and therefore as a consequence of the central-limit theorem it seems reasonable to assume that the sum of their effects will result in a random Hamiltonian with independent Gaussian distributed elements. Such a Hamiltonian will be drawn from one of the Gaussian ensembles of random matrix theory. Recent measurements of the low frequency flux noise in superconducting flux qubits exhibit a coloured noise spectrum \cite{lanting-2009}. We therefore assume that the noise term, $\delta h(\lambda)$, evolves in time as a Ornstein-Uhlenbeck type process, which is a simple example of a random process with a coloured spectrum:
\begin{equation}
	\dot{\delta h}(\lambda)=-\tau \delta h(\lambda)+\epsilon\eta(\lambda)
	\label{eqn:noiseDE}
\end{equation}
where $\tau$ is a correlation time, $\epsilon$ is the noise amplitude, and $\eta(\lambda)$ is a random matrix valued stochastic process where $\left\langle\eta(\lambda)\right\rangle=0$ and $\left\langle\eta(\lambda)\eta(\lambda')\right\rangle=\delta(\lambda-\lambda')$.


\section{The CNOT gate} 
	
	As mentioned previously AQC has been shown to be polynomially equivalent to the circuit model of quantum computing.The `ground state quantum computing' (GSQC) formalism, described in \cite{mizel-2006}, offers the most practical method of constructing a $H_0$ that encodes an arbitrary $M$ qubit, $N$ step quantum circuit. In the GSQC formalism, each of the $M$ qubits in the circuit is viewed as a single electron that can occupy the states in an array of $2\times (N+1)$ quantum dots; where the rows in the array represents either the \ket{0} or \ket{1} states of the qubit. The state of the $m$th qubit during the $n$th step of the algorithm is given by the probability amplitude of the electron being found on the quantum dots denoted by the indices $(m,n,0)$ and $(m,n,1)$. This theoretical construction only incurs a polynomial overheard ($O(N)$) in hardware.
  
  The CNOT gate is one of the simplest entangling quantum gates and when used in conjunction with single qubit rotations it forms a set of universal gates. Therefore, any quantum algorithm can be constructed using a combination of these gates, because of this property we assume that the adiabatic CNOT gate is a representative example of a prototypical adiabatic quantum algorithm. To construct $H_0$ for a CNOT gate using the GSQC formalism, we envisage an array of 8 quantum dots (as shown in the inset of \figref{fig:optimalNoise}), which corresponds to a system of 4 physical qubits. Following the procedure described in \cite{mizel-2006} we can then write;
\begin{multline}
	H_{CNOT}=\left(c^{\dag}_{010}C^{\dag}_{11}-c^{\dag}_{000}C^{\dag}_{10}\right)\left(C_{11}c_{010}-C_{10}c_{000}\right) \\
						+\left(c^{\dag}_{011}C^{\dag}_{11}-c^{\dag}_{001}C^{\dag}_{10}\sigma_{x}\right)\left(C_{11}c_{011}-\sigma_{x}C_{10}c_{001}\right) \\
						+C^{\dag}_{00}C_{00}C^{\dag}_{11}C_{11}+C^{\dag}_{01}C_{01}C^{\dag}_{10}C_{10},
						\label{eqn:CNOTHamiltonian}
\end{multline}
where $c^{\dag}_{mnj}$ is a fermionic creation operator that creates an electron on the corresponding quantum dot, $C^{\dag}_{mn}=\left(c^{\dag}_{mn0},c^{\dag}_{mn1}\right)$ and $\sigma_{x}$ is a Pauli matrix. The ground state energy of the CNOT Hamiltonian \eqref{eqn:CNOTHamiltonian} is zero. To specify the initial state of the qubits before the gate operation we add a small energy penalty to \eqref{eqn:CNOTHamiltonian}; e.g. of the form $H_{Init}=\mu\left(c^{\dag}_{000}c_{000}+c^{\dag}_{100}c_{100}\right)$ for the operation $\ket{00}\rightarrow\ket{00}$. 

	We numerically solve the equations \eqref{eqn:extPYsystem} and \eqref{eqn:noiseDE}. For stability reasons we use the multi-step Adams-Moulton method to solve \eqref{eqn:extPYsystem}. For the sake of generality, the initial conditions for the Pechukas gas, $x_{n}(\lambda=1)$, $v_{n}(\lambda=1)$ and $l_{nm}(\lambda=1)$, are calculated using a perturbation theory expansion in terms of $Z^{-1}$ as at $\lambda=1$, $\mathcal{H}(1)=H_{0}+ZH_{b}+\delta h(1)\equiv Z\left(H_{b}+Z^{-1}H_{0}+Z^{-1}\delta h(1)\right)$. We assume that $\mathcal{H}(1)$ has a non-degenerate, well spaced energy spectrum. Note that the initial conditions contain all the information about the final Hamiltonian $H_{0}$. The initial noise term $\delta h(\lambda=1)$ will be a random matrix drawn from the GOE with amplitude $\epsilon$. Throughout the paper, we take $\mu=-0.1$, $Z^{-1}=0.1$ and $\tau=0.1$.
	

\section{Energy spectra} 
	
\begin{figure}
	\centering
	\includegraphics[width=0.6\columnwidth]{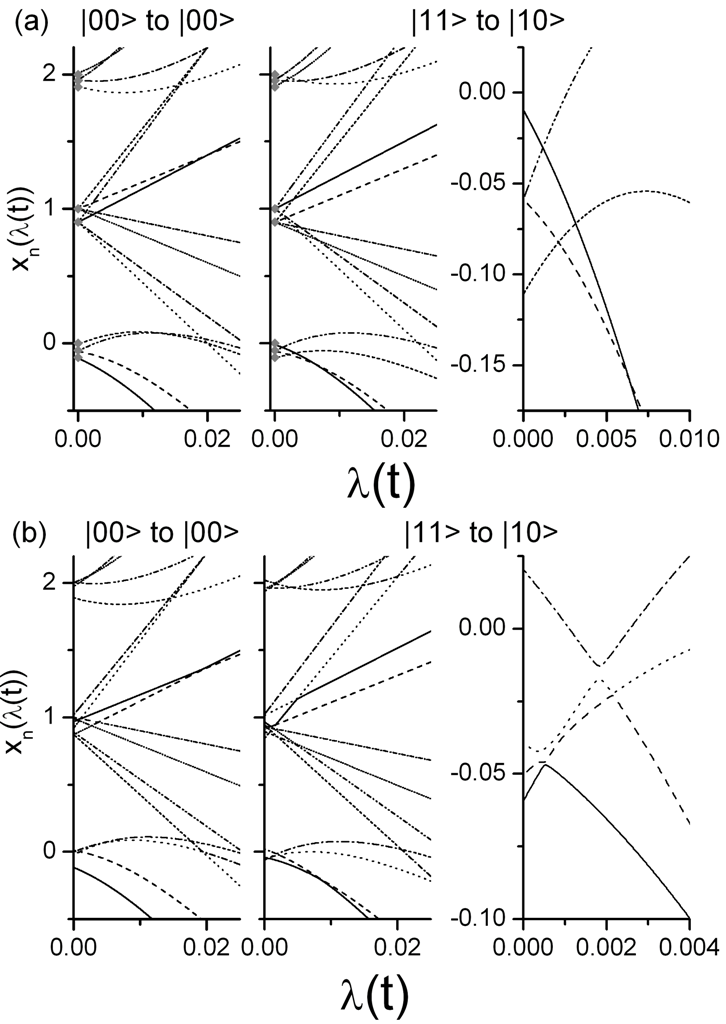}
	\caption{(a) The energy spectra of $\ket{00}\rightarrow\ket{00}$ and $\ket{11}\rightarrow\ket{10}$ operations of the adiabatic CNOT gate in the absence of noise. The solid grey diamonds show the results of direct diagonalisation of the $H_{0}$. (b) Examples of the spectra for $\ket{00}\rightarrow\ket{00}$ and $\ket{11}\rightarrow\ket{10}$ operations of the CNOT gate with a GOE random matrix noise term in the Hamiltonian with amplitude $\epsilon=0.1$. For illustrative purposes, different line styles are used to denote the different energy levels and the third plot in each of the subfigures shows an enlarged view of a region of the $\ket{11}\rightarrow\ket{10}$ spectrum containing a number of crossings.}
	\label{fig:spectra}
\end{figure}

	Figure \ref{fig:spectra}(a) shows the energy spectra of the CNOT gate acting on the \ket{00} and \ket{11} computational basis states. In both cases, the results from the eigenvalue dynamics simulations agree with the results of direct diagonalisation of $H_{0}$ to four significant figures. In both of the spectra, there is an abundance of level crossings which arise because of the symmetries of $\mathcal{H}(\lambda(t))$. 
	
	Degeneracies between the ground and first excited states occur in the $\ket{11}\rightarrow\ket{10}$ operation, which will result in a success probability of zero for this ideal case. This case can be viewed as an example of AQC with an ``improper'' choice of initial configuration, i.e. where $[H_{0},H_{b}]=0$. It is clear that the addition of any type of perturbation (e.g. noise) will break these symmetries, therefore splitting the degeneracies and resulting in a finite success probability. The results of the addition of noise are shown in plot \figref{fig:spectra}(b) and it is evident that there are now no degeneracies in either of the spectra. The spectra for the $\ket{01}\rightarrow\ket{01}$ and $\ket{10}\rightarrow\ket{11}$ operations were found to show similar trends to the $\ket{11}\rightarrow\ket{10}$, in that they also exhibited a degeneracy between the ground and first excited states in the absence of noise.


\section{The effects of noise on level occupation statistics} 
	
	To properly characterise the effects of noise on the adiabatic quantum algorithm for the CNOT gate it is necessary to look at its effects on the probability of level occupation and in particular the success probability of the algorithm (i.e. $P(n=0;\lambda(t)=0|n=0;\lambda(t)=1)$). During the evolution of an adiabatic quantum computer the main mechanism by which the system can tunnel from one state to another is Landau-Zener-St\"{u}ckelberg tunneling \cite{landau-1932,zener-1932,stueckelberg-1932}. This occurs when the separation of two adjacent energy levels is at a local minimum (i.e. at an avoided or level crossing) and the probability of excitation from \ket{n} to \ket{n+1} via this mechanism is given by
\begin{equation}
	P_{LZS}=\exp\left(\frac{-\Delta_{min}^{2}}{\left|\bra{n}ZH_{b}\ket{n+1}\right|\left|\dot{\lambda}(t)\right|}\right)
	\label{eqn:LZS}
\end{equation}
, where $\Delta_{min}$ is the minimum separation between $x_{n}$ and $x_{n+1}$. We assume that the system undergoes uniform evolution (i.e. $|\dot{\lambda}|=\frac{1}{T}$, where $T$ is the computation time) and that the system can initially be found in the ground state with certainty (i.e. $P(n=0;\lambda(t)=1)=1$). Given this information, it is possible to calculate the level occupations as a function of time by identifying all the avoided or level crossings in the spectrum and applying the LZS equation, \eqref{eqn:LZS} to them. 

\begin{figure}
	\centering
		\includegraphics[width=0.7\columnwidth]{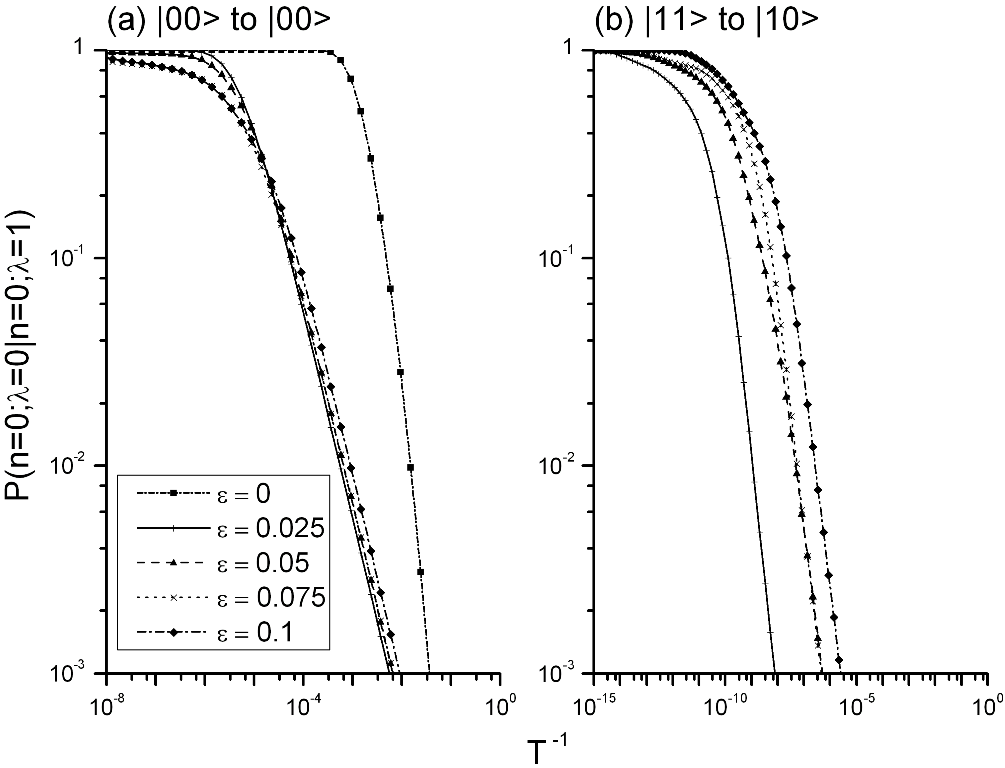}
	\caption{Plot of the success probability, averaged over a number of noise realisations, against computation speed for the $\ket{00}\rightarrow\ket{00}$ and $\ket{11}\rightarrow\ket{10}$ operations of the CNOT gate at various noise amplitudes. Fits to the polynomial regions, i.e. where $P\propto T^{-\gamma}$, of the curves yield exponents of; $\gamma=4/3$ for the $\epsilon=0$ case of the $\ket{00}\rightarrow\ket{00}$ operation, and $\gamma\approx1$ for all other cases.}
	\label{fig:success}
\end{figure}

\begin{figure}
	\centering
		\includegraphics[width=0.65\columnwidth]{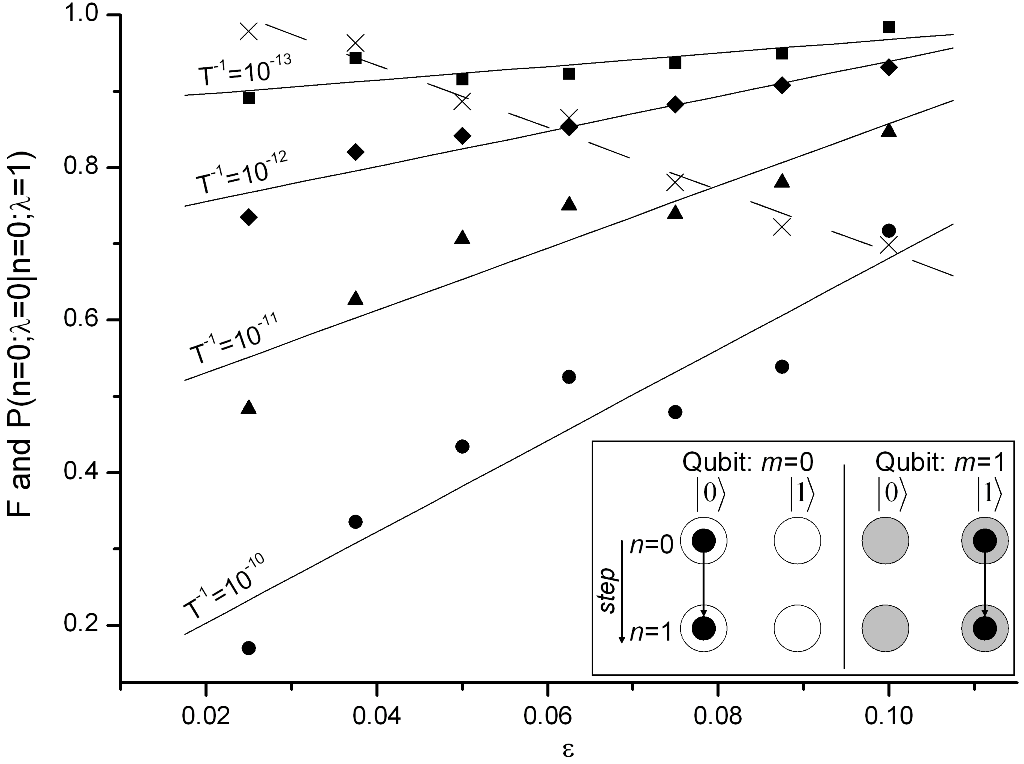}
	\caption{Plot showing the trade-off between the fidelity of the final state (dashed line and crosses) and the average success probability (solid lines and shapes) as functions of noise amplitude for the $\ket{01}\rightarrow\ket{01}$ operation. The inset schematic shows the GSQC representation of the $\ket{01}\rightarrow\ket{01}$ operation of the CNOT gate.}
	\label{fig:optimalNoise}
\end{figure}

	Figure \ref{fig:success} shows the dependence of the average success probability on the computation speed at a range of noise amplitudes for the $\ket{00}\rightarrow\ket{00}$ and $\ket{11}\rightarrow\ket{10}$ operations. In general, the success probability scales polynomially with the speed, i.e. $P\propto T^{-\gamma}$ with $\gamma\approx1$, before approaching unity asymptotically. The exponent is different to those found in \cite{zagoskin-2007} for Hamiltonians drawn from the GUE, however there is no reason to expect a single exponent to hold universally for all choices of $H_0$. From a computational performance point of view, it is important to note that the scaling exponent is independent of the noise amplitude, \i.e. only a prefactor change. This prefactor change could actually be viewed as being beneficial as we can see that the success probability at a given computation speed increases linearly with the noise amplitude (also shown in \figref{fig:optimalNoise}). This occurs as $\Delta_{min}$ increases proportionally with $\epsilon$ and hence this effect is more pronounced for the $\ket{11}\rightarrow\ket{10}$ operation where the existence of the ground state gap is purely due to noise-induced level splitting. As the $\ket{01}\rightarrow\ket{01}$ and $\ket{10}\rightarrow\ket{11}$ operations also have degenerate ground state gaps, their success probability curves behave in a very similar way to the plot shown in \figref{fig:success} for the $\ket{11}\rightarrow\ket{10}$ operation.
	
	The increase in success probability with the noise amplitude (at a specific speed) may not be wholly beneficial though, as it will come at the cost of the fidelity of the final state (i.e. $F=\left|\braket{0_{ideal}(\lambda=0)}{0_{noise}(\lambda=0)}\right|$, where $0\leq F\leq1$). This is because the noise fluctuations will drive the state away from the ideal (i.e. in the absence of noise) path of evolution.  The dependence of the fidelity on noise amplitude is shown in \figref{fig:optimalNoise}. The fidelity of the final state is an important quantity which needs to be maximised to ensure that readout will yield the desired solution. For the three operations where the existence of the ground state gap is due to noise-induced level splitting, there will be a noise amplitude that offers the optimal compromise between the success probability and the fidelity at a given speed and this is shown by the intersections of the curves in \figref{fig:optimalNoise}. These results suggest that the conventional stipulation that $[H_{0},H_{b}]\neq0$ is unnecessary in practical realisations, as when the effects of a generic noise source are taken in to consideration, any hidden symmetries in $\mathcal{H}(\lambda(t))$ will be broken naturally.


\section{Effects of an artificial noise source} 
	
\begin{figure}
	\centering
		\includegraphics[width=0.7\columnwidth]{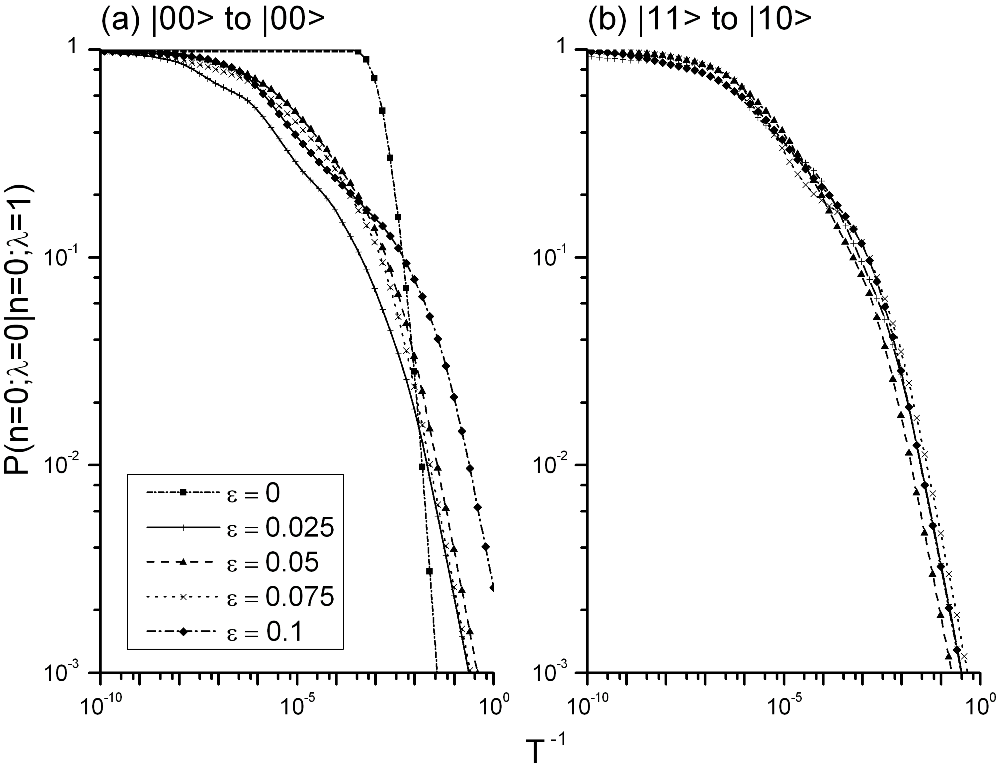}
	\caption{Plot of the success probability, averaged over a number of noise realisations, against computation speed for the $\ket{00}\rightarrow\ket{00}$ and $\ket{11}\rightarrow\ket{10}$ operations of the CNOT gate with an artificial noise source with a time-dependent amplitude and $\alpha=10$. }
	\label{fig:successTD}
\end{figure}

	So far we have viewed $\delta h(\lambda)$ as the natural effect of a number of noise sources that are coupled to the system. If we now envisage a physical system with a negligible level of intrinsic noise. In this situation, it may be beneficial to add an artificial random perturbation to the system. This would be in order to break any degeneracies in the spectrum and offer an alternate, and possibly more efficient, path between the initial and final Hamiltonians. A perturbation term with a time-dependent amplitude, which is large enough to widen the energy gaps at avoided crossings throughout the majority of the computation process but then tends to $0$ as $\lambda(t)\rightarrow0$, would be preferable. For example we could take
\begin{equation}
	\epsilon(\lambda)=\epsilon_{0}\tanh(\alpha \lambda)
	\label{eqn:noiseAmp}
\end{equation}
, where $\alpha$ is a constant determining the rate of decay at $\lambda(t)\gtrsim0$. This would ensure that $\delta h(\lambda)_{ij}$ will not be significantly larger than $H_{0 ij}$ at $\lambda(t)\gtrsim0$, where the bias term is small and the levels are densely packed.  The results of simulations performed using a time-dependent amplitude of the form of \eqref{eqn:noiseAmp} are shown in \figref{fig:successTD}. On average, a given success probability can be achieved at a much faster computation speed, with an improvement of over $10^2$ in some cases. In this idealised situation it is also clear to see that the fidelity of the final solution state would be unaffected (i.e. $F=1$) by this artifical noise signal as $\epsilon(\lambda=0)=0$.


\section{Conclusions} 
	
	We generalise the Pechukas-Yukawa equations to the stochastic case by including an additional noise term in the Hamiltonian. This was used in conjunction with a simple, yet generic, noise model, based on random matrices and a coloured stochastic process, to investigate the effects of noise on the adiabatic algorithm for the CNOT gate. We found that, in general,  the success probability of the algorithm scaled polynomially as a function of computation speed and this scaling was independent of the amplitude of the noise. We demonstrate that when the effects of noise are taken in to account, the criteria used to select an initial configuration for the system may be relaxed and it is not necessary to avoid symmetries in the Hamiltonian. This is because, the presence of noise will break any degeneracies in the energy spectrum, crucially those that exist between the ground and first excited states. In these situations, the success probability at a given computation speed was found to  increase linearly with noise amplitude. It was noted that this increase comes at the expense of the fidelity of the final state, but an optimal compromise between the two factors exists. The effects of an artificial noise source with a time-dependent noise amplitude were also investigated. This scheme was found to offer significantly higher success probabilities at relatively fast computation speeds and could be engineered in such a way that the fidelity of the solution state is unaffected.

%
\end{document}